\documentstyle[11pt,epsfig]{article}
\input epsf.tex
\newcommand{\blankline}{\vskip .3cm}
\def\diffsigma{{\rm Diff}\Sigma}

\def\ldiffm{{\rm LDiff}{\cal M}}
\def\diffm{{\rm Diff}{\cal M}}
\def\cinfty{C^{\infty}(\Sigma)}
\def\product{{{\rm Diff}\Sigma}\odot{C^{\infty}(\Sigma)}}

\def\M{{\cal M}}

\def\ham{{\cal H}_{\perp}}
\def\mom{{\cal H}_i }
\def\momj{{\cal H}_j }

\def\K{{\cal K}}

\def\V{{\cal V}}
\def\beq{\begin{equation}}
\def\eeq{\end{equation}}
\def\bea{\begin{eqnarray}}
\def\eea{\end{eqnarray}}
\def\bes{\begin{displaymath}}
\def\ees{\end{displaymath}}
\def\beas{\begin{eqnarray*}}
\def\eeas{\end{eqnarray*}}
\def\ben{\begin{enumerate}}
\def\een{\end{enumerate}}
\def\adm{{\small ADM }}
\def\kuchar{Kucha\v r }
\begin{document} 
%%%%%%%%%%%%%%%%%%%%%%%%%%%%%%%%%%%%%%%%%%%%%%%%%%%%%%%%
\begin{flushright}
{\small IMPERIAL/TP/96--97/26}
\end{flushright}
\centerline{\large\bf 4D diffeomorphisms in canonical gravity and abelian deformations}
\blankline
\rm
\centerline{Frank Antonsen${}^{*}$ and Fotini Markopoulou}
\blankline
\centerline{\it  Theoretical Physics Group }
\centerline{\it Blackett Laboratory}
\centerline{\it Imperial College of Science, Technology and Medicine}
\centerline{\it London SW7 2BZ}
\blankline
\centerline{February 21, 1997}
\vfill

\centerline{\bf Abstract}
\blankline
A careful study of the induced transformations
on spatial quantities due to 4-dimensional spacetime diffeomorphisms
in the canonical formulation of general relativity is undertaken. 
Use of a general formalism, which indicates the r\^ole of the embedding 
variables in a transparent manner, allows
us to analyse the effect of 4-dimensional
 diffeomorphisms more generally than is possible
in the standard ADM approach. This analysis clearly
indicates the assumptions which are necessary in order
to obtain the ADM--Dirac constraints, and furthermore
shows that there are choices, other than the ADM hamiltonian 
constraint, that one can make for the deformations
 in the ``timelike'' direction. In particular an abelian
generator closely related to true time evolution appears
very naturally in this framework. This generator, its relation
to other abelian scalars discovered recently, and the possibilities
it provides for a group theoretic quantisation of gravity are discussed.  
\vskip 1cm
${}^{*}$ Permanent address: 
Niels Bohr Institute, Blegdamsvej 17, DK 2100 Copenhagen \O,
Denmark\\
\indent email addresses: antonsen@nbivax.nbi.dk, f.markopoulou@ic.ac.uk
\eject

\section{Introduction}
The standard \adm  formulation of canonical general relativity \cite{ADM,Dirac}
may be considered as an initial value problem, defined by
considering initial canonical data on a arbitrary spatial slice in a 
spacetime foliated by a stack of such slices. The spatial
slice, let us call it $\Sigma_t$ (assuming that it is a collection of 
equal-time points), has a 3-metric, $g_{ij}(x)$, inherited from the 
4-metric, $\gamma_{\alpha\beta}(X)$ of the surrounding spacetime $\M$ and 
a momentum $p^{ij}(x)$ conjugate to the metric.
\footnote{
Notation: Greek indices run from 0 to 3, and Latin indices from 1 to 3.
$\Sigma $
denotes 3-dimensional space with coordinates $x^i$, $\Sigma_t$ is an 
equal-time spatial slice, $\M$ denotes 4-dimensional spacetime with 
coordinates $X^\mu$. $\ldiffm$ is
the Lie algebra of 4-dimensional diffeomorphisms $\diffm$. } 
One uses this spatial slice to orient an orthogonal  basis
$(Nn^{\alpha}, N^i X^{\mu}_i )$, defined by the direction  normal to the 
slice, $n^{\mu}$,  and the three tangential directions $X^{\mu}_i\equiv
\frac{\partial X^{\mu}}{\partial x^i}$. 
$N$ and $N^i$ are the lapse and
shift. Any quantity of interest from covariant general relativity is then 
decomposed with respect to this basis. Thus, the canonical theory is 
obtained by decomposing the Hilbert-Einstein action with respect to 
$(Nn^{\mu}, N^i X^{\mu}_i)$. The result describes how the canonical data 
$g_{ij}(x)$ and $p^{ij}(x)$ are  propagated in these four directions by 
four constraints, the hamiltonian constraint $\ham$ in the normal direction 
and the momentum constraints $\mom$ tangentially. 

We shall be specifically concerned with the $\ham,\ \mom$ constraints as 
generators of normal and tangential deformations in the sense described 
above  (as proven in \cite{HKT}). For the canonical representation
 of the Einstein theory, one
also requires the algebra of the constraints to describe the result of 
one deformation followed by another. This is usually referred to as the  
Dirac algebra:
\bea
\left\{\ham(x),\ham(x')\right\}&=&g^{ij}(x)
	\mom(x)\delta_{,j}(x,x')-(x\leftrightarrow x')\\
\left\{\ham(x),\momj(x')\right\}&=&{\cal H}_{\perp,i}(x)\delta(x,x')+\ham(x)
		\delta_{,i}(x,x')\\
\left\{\mom(x),\momj(x')\right\}&=&\momj(x)\delta_{,i}(x,x')
		-(ix\leftrightarrow jx').
\eea

The last line in the Dirac algebra, the Poisson bracket between the 
two momentum constraints, is the statement of $\diffsigma$ invariance on 
$\Sigma_t$: one spatial deformation followed by another is equivalent to 
an overall spatial deformation. The second
line is simply the transformation of $\ham(x)$ as a scalar of density 
of weight 1 
under $\diffsigma$. One needs to be more careful with the first line, the 
Poisson bracket of two hamiltonian constraints. As Hojman, \kuchar and 
Teitelboim explain in \cite{HKT}, this
bracket describes how, if one uses $\ham$ to move from an initial to 
a final slice via an intermediate one, the arrival point on the final slice 
depends on the choice of the intermediate slice. This path-dependence 
of the $\ham$ deformation makes the hamiltonian constraint somewhat
difficult to use, and is responsible for the explicit appearance of the 
metric field $g^{ij}(x)$ in 
the right hand side of the $\left\{\ham(x),\ham(x')\right\}$ Poisson bracket. 

This is a very problematic feature of the Dirac algebra. The metric 
$g^{ij}(x)$ is not a structure constant but one of the fields, 
which means that the Dirac algebra is not a true Lie algebra. 
The existence of powerful group theoretic 
techniques which may be employed in the 
quantisation of theories classically described by Lie algebras
means that the right hand side of this Poisson bracket is unfortunate. 
It stands as an obstacle to any attempt to apply group theoretic 
quantisation methods to the dynamical part of the canonical gravity theory.
\footnote{
The canonical formulation of gravity ought to be particularly convenient 
for a group theoretic approach to quantisation.
 Control over the invariance
group of the theory would enable one to  construct specific,
self-adjoint representations of its Lie algebra, i.e.\ quantum versions
of the constraints and/or canonical variables, acting on an
appropriate Hilbert space \cite{CJI}.
 The kinematical part of such an approach, the canonical commutation 
relations,  has been addressed by Isham and Kakas with promising 
results \cite{IsKa}.  
Unfortunately, but perhaps not surprisingly, the dynamical part, 
including the hamiltonian generators in the scheme, 
has proved a more difficult problem, with central obstacles being
the Dirac algebra  and the hamiltonian constraint. }

In this paper we shall reconsider the Dirac algebra, listing which 
assumptions of the \adm analysis make it unavoidable, and keeping an open 
mind for alternative algebras of generators of deformations in pure gravity. 
The motivation for this work was the discovery by Brown and \kuchar of a
candidate algebra for gravity of the form Abelian$\times\diffsigma$ 
\cite{BrKu}. It was discovered in the context of a non--derivative
coupling of  incoherent dust to gravity. Performing a canonical 
decomposition of the system, they found the surprising result that 
the dust field helped one to select 
a particular scalar combination of the gravitational constraints, a quantity 
consisting purely of gravitational variables which, furthermore, had 
the property of being abelian. More specifically, for incoherent dust, 
this scalar is $G(x):=\ham^2(x)-g^{ij}(x)\mom(x)\momj(x)$ (or rather its 
square root) and satisfies $\{G(x),G(x')\}=0$. 

Brown and \kuchar concluded with a promising proposal. The scalar density 
$G$ is a function of the gravitational variables, like the hamiltonian 
constraint $\ham$, and thus if one used $G$ instead of $\ham$, together with 
the standard $\diffsigma$ constraints, the algebra for gravity would not 
be the problematic Dirac algebra but would instead have the form 
Abelian$\times\diffsigma$. At first sight, it is unclear how this proposal 
can be implemented. For example, $G(x)$ is quadratic in the 
old constraints and hence, if the dust field is removed, it does not 
generate motion on the constraint surface of pure gravity. This 
problem does not arise if the gravitational field remains coupled
to some matter field.  Consequently, 
the dilemma arises as to whether the abelian constraints should be 
investigated in the context of reference fluids and clocks, or 
in pure gravity. The first option avoids the above problem and has 
been investigated in \cite{KuRo, BrMa, IK} who generalised \cite{BrKu} 
to scalar fields and perfect fluids and discovered even more abelian scalar
densities,
which we shall term \kuchar constraints. However, this sidesteps 
the most intriguing feature of these scalars, the fact that they only 
involve gravity variables. 

It has been shown recently \cite{FM} that for pure gravity there is a 
whole family of such abelian scalar densities, including those found 
via particular 
matter couplings, which are solutions of a nonlinear partial 
differential equation. Such a \kuchar scalar density 
$\K[({\rm det}g),\ham,\mom]$ of weight $\omega$ can be 
incorporated in the ``\kuchar algebra''
\bea
\left\{\K(x),\K(x')\right\}&=&0,\\
\left\{\K(x),\mom(x')\right\}&=&\K_{,i}(x)\delta(x,x')+
		\omega\K(x)\delta_{,i}(x,x'),\\
\left\{\mom(x),\momj(x')\right\}&=&\momj(x)\delta_{,i}(x,x')
		-(ix\leftrightarrow jx').
\eea

In the present paper, the discovery of the \kuchar scalars and algebra is 
only the motivation for a search for abelian generators of deformations in 
pure gravity. We will not attempt here to derive the precise form of the 
\kuchar scalars from our results,
although we discuss a possible relationship. Our focus is evolution as an
abelian 
timelike deformation produced by scalar generators. We shall identify how the 
hamiltonian constraint and its algebra is tied to the \adm concept of spatial 
slices and the normal to the slice, which is unrelated to genuine time 
evolution. We find that, if the 3+1 split does not follow the convenient 
route of the orthogonal basis of lapse and shift, one can find scalar 
generators of abelian deformations which have a close relationship 
to time evolution. 

The interesting feature is that the most suitable method for obtaining the 
above results is to consider the long-standing issue of the r\^ole of 
spacetime diffeomorphisms in the canonical theory. We shall discuss 
how spacetime diffeomorphisms can be handled canonically if one takes 
into account the ways in which the space $\Sigma$ is embedded in 
spacetime $\M$ (for globally hyperbolic spacetime, $\M\sim\Sigma\times R$) 
and how $\diffm$ is hidden in the \adm analysis because this embedding is 
treated as fixed. A more suitable picture of spacetime $\M$ as a bundle 
with  fibres $R$ over space $\Sigma$, which naturally accommodates embeddings, 
is proposed in section 2. In section 3, we use this picture to write down 
induced spacetime diffeomorphisms on spatial objects. We then move closer to 
the usual representation of deformations in canonical theory by writing the 
induced spacetime diffeomorphisms as Lie derivatives on tensor quantities, 
for example the 3-metric $g_{ij}(x)$ (section 4). From these general 
transformations, for 
particular choices of diffeomorphisms and embeddings, one can derive the 
usual {\small ADM}-Dirac generators and understand more precisely the 
assumptions that go into the construction of the normal deformation by the 
hamiltonian constraint, as we show in section 5. Interestingly, 
we also find a generator which is in many ways more natural than the 
hamiltonian constraint corresponding to diffeomorphisms along
the $R$ fibre. This constraint is abelian and, in contrast
to the hamiltonian constraint, the evolution it generates
can be more naturally associated with timelike evolution. This particular
choice is discussed in Section 6, and the consequences for
quantisation, along with other concluding remarks are given in
Section 7. 

\section{$\diffm$ and the embedding of $\Sigma$ in $\M$}
The 4-dimensional formulation of general relativity is covariant
under diffeomorphisms ($\diffm$) of the spacetime manifold $\M$. 
In order to develop a Hamiltonian formulation for the purposes of
canonical quantisation one must introduce a 3+1 split of spacetime 
into space and time. While not manifestly covariant, it is clear that 
this representation must still exhibit the symmetries of the 
4-dimensional
theory if only in terms of an arbitrariness of the embedding of the
spatial slice. The actual question of how the 
$\diffm$ covariance is realised in the canonical theory is clearly
of importance. However, the \adm formulation is not necessarily the most
appropriate formalism in which to address this question. While
in 4 dimensions we have the $\ldiffm$ algebra, in the \adm formalism
the only algebraic-like structure is the Dirac algebra. 
It is accepted that the Dirac algebra is, somehow, the ``projection'' of 
$\ldiffm$ onto the foliated spacetime. However, this is not a clear statement. 
The Dirac algebra is very far from being either isomorphic or a subalgebra 
of $\ldiffm$ since it is not even a true algebra. 

Recovering $\diffm$ in the canonical theory is difficult, essentially 
because a fundamental tenet of a canonical theory is {\it not} to have 
explicit reference to what appears as ambient spacetime. Fortunately, 
as has been pointed out in detail by Isham and \kuchar 
\cite{IsKu}, there is indeed a link provided between space and spacetime. 
It is encoded in the way space is thought of as embedded in spacetime in 
a 3+1 theory. That is, in the common assumption of a globally 
hyperbolic spacetime, $\M\sim\Sigma\times R$, there are many ways in 
which $\Sigma$ is embedded in $\M$ (provided 
the metric induced on $\Sigma$ can be spacelike). 
However, in the \adm approach once the 3+1 decomposition is accomplished
one appears to lose contact with details of the embedding. 

In order to carefully analyse the realisation of 4-dimensional symmetries in the
3+1 theory it is clearly necessary to have explicit reference to the
embedding information at the canonical level. For this reason it is 
important to know at which stage of the \adm approach one loses the
explicit embedding information, at least in the sense to which
this information is arbitrary and one can modify the particular 
embedding if required.
 
The procedure of the decomposition is to assume that $\M$ is foliated 
by (equal-time) spacelike slices $\Sigma_t$. If we label coordinates 
in $\M$ by $X^\alpha$ and in $\Sigma_t$ by $x^i$, then the Jacobian 
$X^\mu_i:={\partial X^\mu\over \partial x^i}$ describes 
the way that $\Sigma_t$ is embedded in $\M$.
Each slice $\Sigma_t$ acquires a 3-metric which is the 
projection of the spacetime metric $\gamma_{\alpha\beta}$ on an orthogonal 
basis defined on the slice via the decomposition of the deformation vector 
$\dot X^\mu$ (where the dot denotes differentiation by time):
\beq
\dot X^\mu=N n^\mu+N^i X^\mu_i.
\label{eq:xdot}
\eeq
However, to use this formula in the canonical analysis one needs to treat 
the embedding  $X^\mu_i$ as fixed. For fixed $X^\mu_i$ the 
spacetime $\M$ becomes a particular stack of slices $\Sigma_t$ for increasing 
$t$. This construction is of course general since (\ref{eq:xdot}) holds 
for all $X^\mu_i$ (producing spacelike slices). However if, at the
level of the canonical theory, one 
wishes to see what happens when the embedding changes one needs to return 
to (\ref{eq:xdot}) and perform the analysis more generally. 
Note that otherwise the choice of decomposition has an effect similar to 
the partial ``gauge-fixing'' of a theory where certain invariances of the 
theory, while still present in the sense that the choice of ``gauge''
is arbitrary, become hidden. In this context we no longer have 
$\M\sim\Sigma\times R$ for all possible embeddings $\Sigma\rightarrow\M$, 
but only for a chosen, albeit arbitrary, example. 
As a result, this fixing of the embedding hides the $\diffm$ covariance 
of the theory. 

The \adm construction is based on this assumption of fixing the
embedding and some of its features are natural only in this context. 
Among the basic objects associated with eq.~(\ref{eq:xdot})
are the geometric {\it spatial slice} and its {\it normal} direction, which
naturally lead to the hamiltonian constraint being the generator of 
normal deformations. In a formulation where the embeddings can be modified, 
the spatial slice and its normal will be less fundamental features. 

As the preceding discussion has indicated, in order to describe 
spacetime diffeomorphisms 
we require a canonical split that can accommodate arbitrary embeddings. 
We shall now outline a straightforward formalism of this type 
which relies on the use of the global hyperbolicity requirement 
$\M\sim\Sigma\times R$.
\begin{figure}
\centerline{\mbox{\epsfig{file=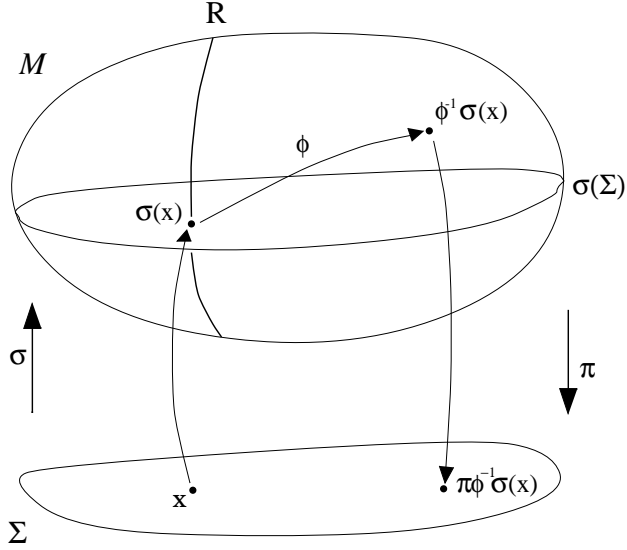}}}
\caption{Spacetime $\M$ as a bundle over space $\Sigma$.} 
\end{figure}
We consider a 3-dimensional manifold $\Sigma$, 
whose metric is {\it not} yet specified. Over each point $x$ of $\Sigma$, 
there is an $R$-line. This results in a line bundle $\M$ over $\Sigma$ 
with fibre $R$:
\beq
	\begin{array}{ccc}
	R&\longrightarrow &\M\\
	&&\pi\downarrow\uparrow\sigma\\
	&&\Sigma
	\end{array}
\eeq
as pictured in figure 1. There is a projection map, $\pi:\M\rightarrow\Sigma$,
and the cross-section map $\pi\circ\sigma=1$. The actual embedding 
then corresponds to this cross-section map $\sigma$, as it takes each 
point $x\in\Sigma$ to a  point $\sigma(x)$ in $\M$. Thus, for every 
$\sigma$ we have an embedding of the 3-dimensional manifold $\Sigma$ in $\M$, 
which we will denote by $\sigma(\Sigma)$. This cross-section 
$\sigma(\Sigma)$ is the spatial slice in the \adm language. 

The bundle $\M$ is $\diffm$ covariant. Under a diffeomorphism $\phi\in\diffm$,
$\sigma(x)\in\M$ is mapped to $\phi^{-1}\sigma(x)\in\M$. The maps 
$\sigma$, and $\pi$ connect $\Sigma$ and $\M$ in a natural way. For example,
when in $\M$, we can act with $\phi\in\diffm$, and finally return to 
$\pi\phi^{-1}\sigma(x)\in\Sigma$ using the projection map $\pi$. As a 
consequence this bundle 
construction allows us to induce spacetime transformations on spatial objects.
The induced spatial transformation is from $x$ to 
$\pi\phi^{-1}\sigma(x)\in\Sigma$. 

In the next two sections, we shall work out explicitly the induced 
$\diffm$ transformations of spatial objects.

\section{Induced spacetime diffeomorphisms on space}
Let us first consider the simplest case, the transformation induced by a 
diffeomorphism $\phi\in\diffm$ on a vector $v_x\in T_x\Sigma$. According 
to figure 1, we can push  this vector forward  through
\beq
T_x\Sigma\stackrel{\sigma_*}{\longrightarrow}T_{\sigma(x)}\M\stackrel
{\phi_*}{\longrightarrow}T_{\phi^{-1}\sigma(x)}\M\stackrel
{\pi_*}{\longrightarrow}T_{\pi\phi^{-1}\sigma(x)}\Sigma
\eeq
sending 
\beq
v_x\in T_x\Sigma\mapsto \sigma_* v_x\mapsto\phi_*\sigma_* v_x 
  \mapsto\pi_*\phi_*\sigma_* v_x\in T_{\pi\phi^{-1}\sigma(x)}\Sigma.
\eeq
In order to evaluate the result we begin with a basis 
$\left({\partial/\partial x^i}\right)_x$ in $T_x\Sigma$ 
with respect to which $v_x$ has components 
\beq
v_x=v^i \left({\partial\over\partial x^i}\right)_x.
\eeq
Then if $\left(\partial/\partial X^\mu\right)_{\sigma(x)}$ is a basis in 
$T_{\sigma(x)}\M$ we can use the Jacobian for the two bases, 
\beq
\sigma^\mu_{,i}(x):=\left(\frac{\partial X^\mu(x)}{\partial x^i}\right)_{\sigma(x)},
\eeq
to obtain the push-forward $\sigma_*v_x$ of $v_x$ as
\beq
\sigma_* v_x\in T_{\sigma(x)}\M = v^i\sigma^\mu_{,i}(x)\left(
{\partial\over\partial X^\mu}\right)_{\sigma(x)}.
\eeq

We now have a vector $\sigma_* v_x$ in $T_{\sigma(x)}\M$ on which we can 
apply a 4-dimensional diffeomorphism $\phi\in\diffm$ and obtain
\beq
\phi_*\sigma_* v_x\in T_{\phi^{-1}\sigma(x)}\M=
v^i\sigma^\mu_{,i}(x)\phi^\nu_{,\mu}\left(\sigma(x)\right)
\left({\partial\over\partial X^\nu}\right)_{\phi^{-1}\sigma(x)}.
\eeq
Finally, we can push this forward to $T\Sigma$ again, using the Jacobian 
$\pi^j_{,\nu}\left(\phi^{-1}\sigma(x)\right)$ for the two bases 
$\left({\partial\over\partial X^\nu}\right)_{\phi^{-1}\sigma(x)}$ and 
$\left({\partial\over\partial x^j}\right)_{\pi\phi^{-1}\sigma(x)}$,
\beq
\pi^j_{,\nu}\left(\phi^{-1}\sigma(x)\right)=
\left(\frac{\partial x^j(X^\nu)}
{\partial X^\nu}\right)_{\phi^{-1}\sigma(x)},
\eeq
to obtain
\beq
\pi_*\phi_*\sigma_* v_x=v^i\sigma^\mu_{,i}(x)\phi^\nu_{,\mu}
 \left(\sigma(x)\right)
\pi^j_{,\nu}\left(\phi^{-1}\sigma(x)\right)
\left({\partial\over\partial x^j}\right)_{\pi\phi^{-1}\sigma(x)}.
\label{eq:pi}
\eeq
Combining the results above, the induced spacetime diffeomorphism 
on the spatial vector $v_x$ has the component form
\beq
v^i_x\mapsto v'^j_{\pi\phi^{-1}\sigma(x)}=v^i
\sigma^\mu_{,i}(x)\phi^\nu_{,\mu}\left(\sigma(x)\right)
\pi^j_{,\nu}\left(\phi^{-1}\sigma(x)\right).
\label{eq:vector}
\eeq

This equation may be readily extended to a spatial vector field. 
This is because, when $x$ is varied smoothly and continuously over all 
$\Sigma$ in (\ref{eq:vector}), this transformation remains well-defined. 
It can therefore be used to push-forward vector fields.
\footnote{Note that 
this mapping of the vector field can not be factorised, as the 
push-forward with $\pi_*$ in (\ref{eq:pi}) is a many-to-one map and 
not defined for a vector field.}

Let us now turn to 1-forms and covectors. In this case it is 
easier to write down the induced $\diffm$ pullback if we reverse the route
used in the analysis of vectors. In other respects the derivation is 
very similar to that above. The component result of the pullback of the 
one-form $\omega(\pi\phi^{-1}\sigma(x))$ to $\omega'(x)$ via 
$\sigma^*\phi^*\pi^*$ is
\beq
\omega'_j(x)\in T_x^*\Sigma=\omega_j\left(\pi\phi^{-1}\sigma(x)\right)
 \pi^j_{,\nu}
\left(\phi^{-1}\sigma(x)\right)\phi^\nu_{,\mu}\left(\sigma(x)\right)
\sigma^\mu_{,i}(x)\left(dx^i\right)_x.
\label{eq:oneform}
\eeq
Similarly, the covector transformation 
$k'\in T^*_{\pi\phi^{-1}\sigma(x)}\Sigma 
\rightarrow k\in T^*_x\Sigma$ is
\beq
k_j(x)=k'_j\pi^j_{,\nu}\left(\phi\sigma(x)\right)\phi^\nu_\mu
\left(\sigma(x)\right)\sigma^\mu_i(x)\left(dx^i\right)_x.
\eeq
One can check that $k$ and $v$, as given by the formulae above, 
are indeed dual i.e. 
$\langle k,\pi_*\phi_*\sigma_* v\rangle_{\pi\phi^{-1}\sigma(x)}=
\langle \sigma^*\phi^*\pi^* k, v\rangle_x$.

\section{Infinitesimal spacetime diffeomorphisms on spatial tensors}

At this stage, we have coordinate expressions for the transformations 
of the simplest tensorial objects and it is straightforward to 
extend these results to other spatial objects as required. Let us now 
return to our initial problem, the relation between $\diffm$ and the 
deformations generated by constraints. We would like to compare the 
present formalism to the standard approach of constraint generators
decomposed with respect to a fixed orthogonal basis. For 
example, we can consider the tangential deformation of the 
3-metric by the momentum constraint 
$\mom$ (smeared by a vector field $N$):
\beq
\left\{{\cal H}(N), g_{ij}\right\}=\delta g_{ij}={\cal L}_N g_{ij}.
\label{eq:mom}
\eeq
We need to work with infinitesimal $\phi\in\diffm$, namely, Lie 
derivatives with respect to a vector field $\V\in T\M$. Such an infinitesimal 
diffeomorphism transforms, say, the covector $\xi\in T^*\M$ in the manner
\beq
\xi\mapsto\xi'=\xi+\epsilon{\cal L}_\V\xi+{\cal O}\left(\epsilon^2\right).
\eeq

Recall that the base space $\Sigma$ does not have a fixed 3-metric 
$g_{ij}(x)$, unlike a spatial slice $\Sigma_t$. 
Instead, for  $\Sigma$, $g_{ij}(x)$
is a special symmetric 2-index tensor, an element of 
$T^*_x\Sigma\otimes T^*_x\Sigma$.
Its deformation (\ref{eq:mom}) will then be a particular induced 
$\diffm$ map $\sigma_*\phi_*\pi_*:T^*_x\Sigma\otimes T^*_x\Sigma\rightarrow 
T^*_{\pi\phi^{-1}\sigma(x)}\Sigma\otimes T^*_{\pi\phi^{-1}\sigma(x)}\Sigma$, 
as we shall verify in section 5.  

In preparation let us write down the induced spacetime diffeomorphism of a 
general tensor in $T^*_x\Sigma\otimes T^*_x\Sigma$, say $t_{ij}(x)$. 
The result follows in a similar manner to the calculations already
presented, except that transformations are required for each index
and we consider only infinitesimal diffeomorphisms with parameter
$\epsilon$, i.e.
\beq
t'_{ij}=t_{ij}+\epsilon\sigma^\mu_{,i}\sigma^\nu_{,j}
 \left[\V^\lambda t_{\mu\nu}
+\left(\partial_\mu\V^\lambda\right)t_{\lambda\nu}+\left(\partial_\nu
\V^\lambda\right) t_{\lambda\mu}\right],
\label{eq:tij}
\eeq
where $t_{\mu\nu}$ (at $\sigma(x)$)  is shorthand for $t_{ij}(x)$ embedded in 
$\M$:\footnote{
For clarity we will ommit some indices. In detail, the transformation (\ref{eq:tij}) is:
\bea
& & t'_{ij}\left(\pi\phi^{-1}\sigma(x)\right)=
t_{ij}(x)+\epsilon\sigma^\mu_{,i}(x)\sigma^\nu_{,j}(x)\nonumber\\
& & \left[\V^\lambda\left(\sigma(x)\right) t_{\mu\nu}\left(\sigma(x)\right)+
\left(\partial_\mu\V^\lambda\left(\sigma(x)\right)
\right)t_{\lambda\nu}\left(\sigma(x)\right)  +\left(\partial_\nu
\V^\lambda\left(\sigma(x)\right) \right) t_{\lambda\mu}\left(\sigma(x)\right)\right]\nonumber.
\eea
In what follows, we will use a prime to denote the value of the tensor at 
point $\pi\phi^{-1}\sigma(x)$. }
\beq
t_{ij}(x)=\sigma^\mu_{,i}(x)\sigma^\nu_{,j}(x) 
t_{\mu\nu}\left(\sigma(x)\right).  
\eeq
Similarly, for a contravariant 2-tensor, $l^{ij}\in T\Sigma\otimes T\Sigma$, 
we have
\beq
l'^{ij}=l^{ij}+\epsilon\pi^i_{,\mu}\pi^j_{,\nu}\left[\V^\lambda l^{\mu\nu}
+\left(\partial_\mu\V^\lambda\right)l^{\lambda\nu}+\left(\partial_\nu
\V^\lambda\right) l^{\lambda\mu}\right].
\label{eq:lij}
\eeq

The transformations (\ref{eq:tij}) and (\ref{eq:lij}) are general
formulae that encode the induced action of arbitrary 
4-dimensional infinitesimal 
diffeomorphisms on spatial 2-tensors. The compactness of these
expressions hides the fact that most of the physical
information is contained in the sets of $\sigma^\mu_{,i}$ and the 
choice of the vector field $\V^\mu$. Recall that
$\sigma^\mu_{,i}$ are the coordinate expressions for 
the embedding $T\Sigma\rightarrow T\M$ induced by the cross-sections
$\sigma:\Sigma\rightarrow\M$. The choice of the vector field $\V^\mu$ 
is determined by the spacetime diffeomorphism $\phi\in\diffm$ we are 
performing. 

In the next two sections we show
that, as special cases of (\ref{eq:tij}) and (\ref{eq:lij}), we can, firstly 
retrieve the Dirac algebra explicitly as an orthogonal projection of 
spacetime diffeomorphisms on $\Sigma\times R$ and, secondly,
obtain abelian transformations generated by an extra class of
diffeomorphisms along the $R$-fibre. These arise very naturally, 
are by construction abelian, and suggest intriguing connections to 
existing 3+1 work. 

\section{The ADM-Dirac generators as projections of 
$\ldiffm$ on an orthogonal basis}

Having developed a formalism for considering the transformation of
spatial tensors under 4-dimensional diffeomorphisms of the
bundle $\M$ that explicitly involves
``embeddings'', we may use it for the Dirac algebra of the canonical
constraints.  Appropriate conditions on the vector field $\V$ via which
the Lie derivatives of the transformations (\ref{eq:tij}) and
(\ref{eq:lij}) are defined and the embedding $\sigma$ will 
reproduce the hamiltonian and
momentum constraints as generators of spatial and  
normal diffeomorphisms.

We begin by using (\ref{eq:tij}) to derive the known deformations of 
the 3-metric $g_{ij}(x)$ under the momentum and hamiltonian 
constraints \cite{KVK}. Recall that for the purpose of considering the
effect of spacetime diffeomorphisms $g_{ij}(x)$ may be regarded as a  
tensor of the form $t_{ij}(x)\in T^*\Sigma\otimes T^*\Sigma$. That is, 
its transformation under a general infinitesimal spacetime diffeomorphism 
is given by equation (\ref{eq:tij}),
\beq
g'_{ij}=g_{ij}+\epsilon\sigma^\mu_{,i}\sigma^\nu_{,j}
\left[\V^\lambda g_{\mu\nu}
+\left(\partial_\mu\V^\lambda\right)g_{\lambda\nu}+\left(\partial_\nu
\V^\lambda\right) g_{\lambda\mu}\right],
\label{eq:gij}
\eeq
with $g_{\mu\nu}(\sigma(x))$ given by 
$g_{ij}(x)=\sigma^\mu_{,i}(x)\sigma^\nu_{,j}(x) g_{\mu\nu}(\sigma(x))$.  
The constraints are then generators of canonical transformations 
between elements of $T^*\Sigma\otimes T^*\Sigma$. 

A spatial diffeomorphism is generated by a vector field $N$ which is 
purely spatial, $N\in T\Sigma$. When $\Sigma$ is embedded in $\M$, the 
corresponding spacetime diffeomorphism will be with respect to a 
vector field $\V$ which lies in the cross-section 
$\sigma(\Sigma)$, i.e. $\V^\mu(\sigma(x))=\sigma^\mu_{,i}(x)N^i(x)$. Using 
the identity $\pi\circ\sigma=1$, namely,
\beq
\pi^i_{,\nu}\left(\sigma(x)\right)\sigma^\nu_{,j}(x)=\delta^i_j(x),
\label{eq:pisigma}
\eeq
we obtain
\beq
\pi^i_{,\mu}\left(\partial_k\sigma^\mu_{,j}\right)=-
\sigma^\mu_{,j}\left(\partial_k\pi^i_{,\mu}\right),
\eeq
which, together with the integrability condition 
\beq
\partial_j\sigma^\mu_{,i}=\partial_i\sigma^\mu_{,j}
\eeq
leads to equation (\ref{eq:gij}) reducing to the expected form:
\beq
g'_{ij}(\pi\phi^{-1}\sigma(x)) = g_{ij}(x)+\epsilon{\cal L}_Ng_{ij}(x)
\label{eq:spatial}
\eeq
as in equation (\ref{eq:mom}).
Therefore, this induced diffeomorphism $g_{ij}\rightarrow g'_{ij}$ is indeed
an element of $\diffsigma$.

Let us now check whether, for $\phi$ a diffeomorphism with respect to a 
vector field normal to the cross-section $\sigma(\Sigma)$, equation 
(\ref{eq:gij}) reduces to the known normal deformation of the 3-metric 
generated by the hamiltonian constraint \cite{JY},
\beq
g'_{ij}=g_{ij}+\epsilon\left[\partial_0 g_{ij}+D_{(i}N_{j)}\right],
\label{eq:ham}
\eeq
where $D_i$ denotes the spatial covariant derivative.
The following derivation is interesting mainly because it shows which 
are the assumptions of \adm needed to make the hamiltonian constraint and 
normal deformations a convenient tool to use.
\footnote{
Of course, in the \adm philosophy the hamiltonian constraint is perfectly 
reasonable, as the normal can be defined intrinsically to the slice and as 
a result one can use quantities such as the extrinsic curvature to
conveniently describe this constraint, and obtain a compact
formulation of the initial-value formulation of 
general relativity. However, in the present context where embeddings play 
an essential r\^ole, the spatial slice is no longer such a central object.}
Note that in our picture of 3-space arbitrarily embedded in spacetime, the 
normal is no longer the most natural direction to use in order
to describe deformations which are not tangential to the embedded slice,
as we shall come to in Section 6. 

It turns out that there are four assumptions used in the \adm 
formulation in order to turn an arbitrary normal deformation, 
i.e. equation (\ref{eq:gij}) for $\V^\mu$ some normal vector field $n^\mu$:
\beq
g'_{ij}=g_{ij}+\epsilon\sigma^\mu_i\sigma^\nu_j\left[ n^\lambda g_{\mu\nu}
+\left(\partial_\mu n^\lambda\right)g_{\lambda\nu}+\left(\partial_\nu
n^\lambda\right) g_{\lambda\mu}\right],
\label{eq:ngij}
\eeq
into the simplified form of (\ref{eq:ham}). Firstly, one needs to choose 
(and fix) the embedding $\sigma$. Once the embedding is fixed, as a
second step, the 
lapse and shift can be introduced, in a manner formally equivalent
to the usual decomposition of the deformation vector
\beq
\dot X^\mu=n^\mu N+X^\mu_i N^i.
\label{eq:def}
\eeq
In our notation $X^\mu_i=\frac{\partial X^\mu}{\partial x^i}
\equiv\sigma^\mu_{,i}(x)$ and $X^\mu\equiv\sigma^\mu(x)$, so the lapse and 
shift will appear through $\partial_0\sigma^\mu$. Explicitly, and similarly 
to the case of spatial diffeomorphisms, we can impose integrability to 
find that
\beq
\partial_i\left(\partial_0\sigma^\mu\right)=\partial_0 \sigma^\mu_{,i},
\label{eq:int}
\eeq
which may be decomposed in the same basis as (\ref{eq:def}) to give
\beq
\partial_0\sigma^\mu_i=\partial_i\left(n^\mu N+\sigma^\mu_{,i} N^i\right).
\label{eq:split}
\eeq
The general normal diffeomorphism (\ref{eq:ngij}) can be simplified to 
\beq
g'_{ij}=g_{ij}+\epsilon n^\lambda\partial_\lambda g_{ab}+\epsilon n^\lambda
\sigma^\mu_{,i}\sigma^\nu_{,j}\left[\partial_\lambda\left(
\pi^a_{,\mu}\pi^b_{\nu}\right)-\partial_\mu\left(\pi^a_{,\lambda}
\pi^b_{,\nu}\right)-\partial_\nu\left(\pi^a_{,\lambda}\pi^b_{,\mu}\right)
\right]g_{ab}.
\eeq
Using the integrability condition (\ref{eq:int}), 
and the decomposition (\ref{eq:split}) we find, after some tedious 
calculations,  
\bea
g'_{ij}&=&g_{ij}+\epsilon n^k\left\{\partial_k g_{ij}-\left(
\sigma^\nu_{,j}\delta^a_{[k}\partial_{i]}\pi^b_{,\nu}+
\sigma^\mu_i\delta_{[k}\partial_{j]}\pi^a_{\mu}\right)
g_{ab}\right\}-\nonumber\\
& & \epsilon n^0\left\{ D_{(i}N_{j)}+\partial_0 g_{ij}-N^k
\left[ \sigma^\mu_{,j}\partial_k\left(\pi^a_{,\mu}g_{ia}\right)+
\sigma^\mu_{,i}\partial_k\left(\pi^a_{,\mu}g_{aj}\right)\right]\right\} - 
\nonumber\\
& & \left[\delta^a_i\pi^b_{,\mu}\partial_j(Nn^\mu)+
\delta^b_j\pi^a_{,\mu}\partial_i(Nn^\mu)\right] g_{ab}+\nonumber\\
& &\left(\partial_{(i}\sigma^\mu_{j)}\right)g_{0\mu}-g_{b(i}\partial_{j)}
\pi^b_0.
\label{eq:normal}
\eea

Requiring that the above transformation $g_{ij}\rightarrow g'_{ij}$ be 
produced by a generator ${\cal F}(g_{ij},p^{ij})$ via $\delta g_{ij}=
\{g_{ij}, {\cal F}(N)\}$ (by analogy to the usual normal transformation 
(\ref{eq:ham}) also being the result of the Poisson bracket of the metric
with the smeared hamiltonian constraint $\{g_{ij},\ham(N)\}$) we can find 
${\cal F}$:
\beq
{\cal F}\left(g_{ij},p^{i},\sigma^\mu_{,i}\right)=
n^0\ham+p^{ij}n^k\partial_k g_{ij}+
p^{ij}A^{ab}_{i}g_{ab}+f(g,\sigma),
\label{eq:f}
\eeq
where $\ham$ denotes the standard normal deformation as in 
eq.\ (\ref{eq:ham}), 
$p^{ij}$ has been defined to be the time derivative of $g_{ij}$,  and 
the other terms are simply the rest of (\ref{eq:normal})
expressed in a convenient notation. $A^{ab}_{ij}$ is the function of lapse, 
shift and embedding
\footnote{The notation $X_{(a|b|c)}$ means that $b$ is not to be included 
in the symmetrisation which then takes place only in $a,\ c$.}
\bea
\label{Aabij}
A^{ab}_{ij}&=&-n^k\left[\sigma^\nu_{,j}\delta^a_{[k}\partial_{i]}\pi^b_{,\nu}
+\sigma^\mu_i\delta^b_{[k}\partial_{j]}\pi^a_{,\mu}\right]+\nonumber\\
& & n^0\left[\delta^a_{(i}\partial_{j)}\pi^b_{,0}-\pi^a_{,0}\pi^b_{,\mu}
\partial_i\sigma^\mu_{,j}+\pi^a_{,\mu}\pi^b_{,0}\partial_j\sigma^\mu_{,i}+
 \right.
\nonumber\\
& & \left. N^k\sigma^\mu_{,(i}\partial_{|k}\pi^a_{,\mu|}\delta^b_{i)}-
\delta^a_i\pi^b_{,\mu}\partial_j(Nn^\mu)-\delta^b_j\pi^a_{,\mu}
\partial_i(Nn^\mu)\right],
\eea
and $f(g,\sigma)$ is an unspecified function of the 3-metric and the 
embedding only. 

The generator ${\cal F}$ in (\ref{eq:f}) is still cumbersome 
because we are only halfway through imposing the \adm assumptions. 
As the third step we now ``lock'' the coordinate frame to 
our embedding choice, so that $n^\mu$ becomes $n^0=-1,\ n^k=0$. 
The second term in (\ref{eq:f}) and the first term in (\ref{Aabij})
then vanish. Finally, let us assume that the cross-sections are 
slices of constant coordinate time which implies 
that $\sigma^\mu_{,i}=const$. The last two terms of (\ref{eq:f})
then vanish as they contain derivatives of $\sigma^\mu_{,i}$ and we 
have recovered the \adm hamiltonian constraint $\ham$. 

\section{Abelian diffeomorphisms along the $R$-fibre}

The derivation in the previous section clarifies the statement that the 
Dirac algebra is the ``projection'' of $\ldiffm$. However, 
this projection is with respect to a basis determined by a spatial slice 
and its normal direction, 
rather than on $\Sigma\times R$. 
In fact, the projection on 
$\Sigma\times R$, which we are now going to consider, 
remarkably leads to a 
generator algebra of the form Abelian$\times\diffsigma$.

\begin{figure}
\centerline{\mbox{\epsfig{file=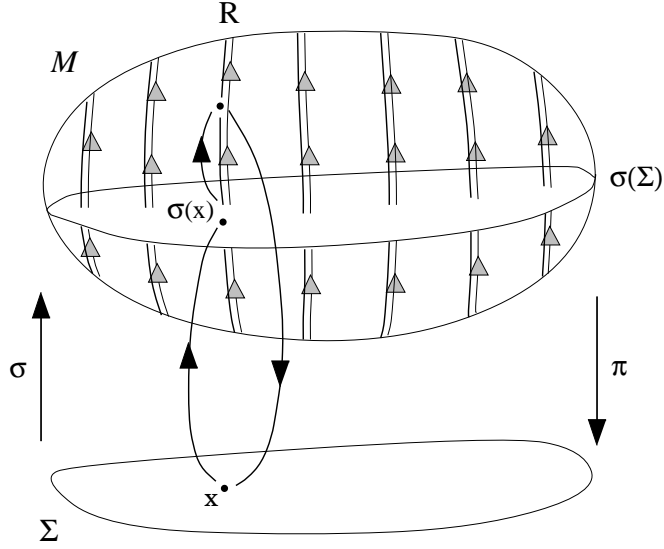}}}
\caption{Abelian deformations.}
\end{figure}

As much as the normal diffeomorphisms were unnatural and rather
tedious to recover, this third special class of diffeomorphisms is 
simple and straightforward to find. It is the case where the spacetime 
diffeomorphism is a base-point preserving map in the bundle. That is, the 
vector field $\V^\mu$ is along the 1-dimensional $R$-fibre, as shown in 
figure 2. By construction, this $\V^\mu$ may be represented as  
$\V=\frac{\partial}{\partial\tau}$, $\tau$ being the affine parameter 
along the fibre. In this case, the transformation (\ref{eq:gij}) 
of the 3-metric reduces to
\beq
g'_{ij}=g_{ij}+\epsilon\frac{\partial}{\partial\tau}g_{ij}-\epsilon\left(g_{kl}
\pi^k_{,\mu}\pi^l_{,\nu}\right)\frac{\partial}{\partial\tau}
\left(\sigma^\mu_{,i}\sigma^\nu_{,j}\right).
\label{eq:abelian}
\eeq
This describes the change in the value of $g_{ij}(x)$ at each point 
$x\in\Sigma$ after some ``time evolution'' $\tau$. Note that the first 
two terms of (\ref{eq:abelian}) reflect this time-evolution property of 
the base-point preserving diffeomorphisms in a straightforward way. 
The third term depends only on the embedding, which changes 
in $\tau$-time since it is not restricted to being static in this formalism. 
Furthermore, because of the simplicity of the spacetime we are dealing with, 
it is not unexpected that this transformation along the 1-dimensional fibre 
is abelian. More accurately, 
our natural assumption that the fibre is an $R$-group acting freely 
on $\M$ lets us treat $\M$ as a principal $R$-bundle. Then the 
above transformations from $\M$ to itself form a group, 
the automorphism group of $\M$, Aut$(\M)$. Moreover, since
$\M$ is trivial, Aut$(\M)$ is isomorphic to the group 
$C^\infty(\Sigma,R)$ of functions on $\Sigma$, which is abelian. Thus 
we have obtained a framework in which the evolution of the embedded
slices is naturally described by abelian constraints.

The result is that this projection of $\diffm$ on $\Sigma\times R$ leads to 
the Lie algebra $L\product$ (with the symbol $\odot$ denoting 
the semidirect product).
One may choose to use the transformation (\ref{eq:abelian}) in place of 
the normal (\ref{eq:ham}) combined with the spatial diffeomorphisms 
(\ref{eq:spatial}) for a 3+1 decomposition with a true Lie algebra of its 
deformation generators. 

The {\small ADM}-Dirac algebra and this $\product$ algebra are very 
special cases of the general spacetime deformations (\ref{eq:tij}) and 
(\ref{eq:lij}) in that they only refer to the 3-space. 
The {\small ADM}-Dirac algebra is constructed from the beginning in this way, 
starting from a spatial slice and using quantities that can be defined 
intrinsically to the slice. The $\product$ algebra also turns out to have 
this property as both $\diffsigma$ and more importantly $\cinfty$ require 
only $\Sigma$ and not $\M$ for
their definition. In fact, it is possible to derive the results of 
this paper without reference to spacetime $\M$ as a physical manifold 
with 4-metric $\gamma_{\alpha\beta}$, but by starting from a 3-dimensional 
space $\Sigma$ on which $\diffsigma$ and $\cinfty$ can be defined. In 
that context, the 4-dimensional bundle is only a helpful way to unfold
transformations under these two groups by raising an $R$-fibre over 
each spatial point and constructing a 
$\Sigma\times R$ bundle over $\Sigma$. 
This approach was followed in \cite{AnMa}. 
One should note that information about spacetime and the spacetime 
metric is not used until the very last stages of the derivation of the 
hamiltonian generator, when ``locking'' the coordinate frame to 
the chosen foliation. 

\section{Conclusions}
Motivated by the recent discovery of abelian constraints, and the
proposal that these abelian generators could be of use
in group theoretic approaches to canonical quantisation \cite{BrKu}, we 
re--analysed the {\small ADM}-Dirac algebra and the hamiltonian 
constraint. We traced the problem of its non-closing Poisson bracket to the 
selection of a spatial slice and its normal as primary elements of 
the \adm canonical analysis and the fixed choice of embeddings
needed for their use. Allowing variation of the embeddings, 
which is in principle allowed in the canonical gravity, makes it 
possible to describe the effect of 4-dimensional spacetime
diffeomorphisms, at least when spacetime is globally hyperbolic. 
In order to include the embeddings, we found it
necessary to change our viewpoint of spacetime from a fixed stack-of-slices 
to spacetime as an $\M\sim\Sigma\times R$ bundle over a generic 
3-manifold $\Sigma$, where 
the embeddings correspond to the cross-section maps from $\Sigma$ to $\M$.
By including embeddings explicitly in this manner we were
able to break $\diffm$ 
covariance in a controlled manner in order to obtain the 
induced $\diffm$ action on spatial quantities. 

Using these general transformations, we were able to perform 
3+1 splits of $\diffm$ corresponding to two different embeddings. A
standard normal and tangential split with respect to the spatial
slice which leads to the {\small ADM}-Dirac algebra, and one on 
$\Sigma\times R$, leading to an Abelian$\times \diffsigma$ algebra. 
The first case was useful in clarifying the {\small ADM} 
assumptions used in the construction of the hamiltonian constraint 
and showing how they are incompatible with truly variable embeddings.  
The second split makes use of the $R$ in $\Sigma\times R$, and perhaps not 
surprisingly, produces abelian deformations whose form resembles 
time evolution (although we have left open the issue of the r\^ole 
of the $R$-fibres). It is important to note that this $L\product$ algebra 
only refers to the space $\Sigma$. Corresponding to the way in which 
the \adm analysis can be thought of as a foliation of spacetime 
$\M$ by spatial slices, the decomposition in terms of the bundle is a 
fibering of $\M$ by $R$. 

While the existence of an abelian algebra for canonical
general relativity is very promising, particularly in the context of group
quantisation, there are a number of tasks which need to be performed before 
deciding whether the Abelian$\times\diffsigma$ split is of practical use. 
So far, we have used Lie derivatives to describe
infinitesimal transormations. In the sense described in \cite{HKT} this 
amounts to constructing the kinematics of the Abelian$\times\diffsigma$
formulation. The dynamics would describe the change of functionals of
the canonical variables $g_{ij}(x)$ and $p^{ij}(x)$ under these 
transformations using Poisson bracket relationships. This requires 
finding expressions for our abelian generators in terms of the 
canonical variables. To find them, one may use the same postulate as 
\cite{HKT} and ask that they should ``represent the kinematical 
generators'', that is, they should be constructed from the canonical 
variables---and the embedding variables in our case---in such a way that 
their Poisson brackets close like the commutators of the corresponding 
kinematical generators. We expect the inclusion of the embedding
variables to produce interesting results \cite{BoMa} and possible 
relationship to \cite{IsKu} and \cite{BeKo}. 
 
Finally, let us recall the \kuchar scalars. In their  derivation  
in \cite{BrKu,KuRo,BrMa} a reference fluid is 
required to select the particular form of the 
scalar, and according to \cite{IK}, each member of the family found in 
\cite{FM} also corresponds to a particular choice of reference fluid. 
The unsatisfactory element there is that, thus far, each case may only
be obtained in a somewhat ad hoc manner.
It appears possible to set up an equivalence between the 
$\sigma$ variables and reference fluids \cite{RB}, thus providing a 
better organised derivation of \kuchar scalars and a connection 
of the present work to the reference fluid results. 

\section*{Acknowledgements}

We are grateful to Chris Isham for his guidance throughout this work. 
Numerous discussions with Julian Barbour, Roumen Borissov, 
Jonathan Halliwell, Adam Ritz and Lee Smolin have been very important 
to our understanding of issues of canonical gravity. 
 FM would like to thank Abhay Ashtekar for hospitality at Penn State 
where this work was completed
and she is partly supported by the A S Onassis
Foundation.

%-----------------------------------------------------------------

\end{document}